\documentclass[twoside,a4paper,11pt]{sea9}
\usepackage{graphicx}
\usepackage{hyperref}
\usepackage{movie15}
\begin{document}
\pagenumbering{arabic}
\pagestyle{myheadings}
\thispagestyle{empty}
{\flushleft\includegraphics[width=\textwidth,bb=58 650 590 680]{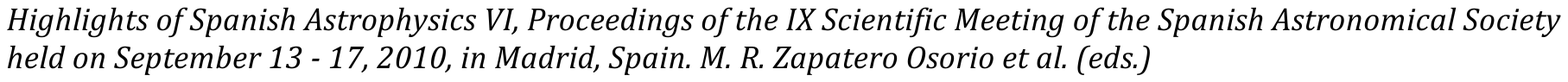}}
\vspace*{0.2cm}
\begin{flushleft}
{\bf {\LARGE
%
Internal kinematic and physical properties in a BCD galaxy: Haro 15 in detail. 
%
}\\
\vspace*{1cm}
%
Ver\'onica Firpo$^{1}$,
Guillermo Bosch$^{1}$, 
Guillermo F. H\"agele$^{2}$,
\'Angeles I. D\'{\i}az$^{2}$,
and 
Nidia Morrell$^{3}$
%
}\\
\vspace*{0.5cm}
%
$^{1}$
Facultad de Ciencias Astron\'omicas y Geof\'{\i}sicas, Universidad Nacional de la La Plata, Paseo del Bosque s/n, 1900 La Plata, Argentina.\\
$^{2}$
Departamento de F\'{\i}sica Te\'orica, C-XI, Univerdidad Aut\'onoma de
Madrid, 28049 Madrid, Spain.\\
$^{3}$
Las Campanas Observatory, Carnegie Observatories, Casilla 601, La Serena, Chile.
%
\end{flushleft}
%
\markboth{
Haro 15 in detail
}{ 
%
Firpo et al. 
%
}
\thispagestyle{empty}
\vspace*{0.4cm}
\begin{minipage}[l]{0.09\textwidth}
\ 
\end{minipage}
\begin{minipage}[r]{0.9\textwidth}
\vspace{1cm}
\section*{Abstract}{\small
%
We present a detailed study of the kinematic and physical properties of the ionized gas in multiple knots of the blue compact dwarf galaxy Haro 15. Using echelle and long slit spectroscopy data, obtained with different instruments at Las Campanas Observatory, we study the internal kinematic and physical conditions (electron density and temperature), ionic and total chemical abundances of several atoms, reddening and ionization structure. Applying direct and empirical methods for abundance determination, we perform a comparative analysis between these regions and in their different components. On the other hand, our echelle spectra show complex kinematics in several conspicuous knots within the galaxy. To perform an in-depth 2D spectroscopic study we complete this work with high spatial and spectral resolution spectroscopy using the Integral Field Unit mode on the Gemini Multi-Object Spectrograph instrument at the Gemini South telescope. With these data we are able to resolve the complex kinematical structure within star forming knots in Haro 15 galaxy. 
%
\normalsize}
\end{minipage}
%
%
%
\section{Introduction \label{intro}}

The aim of this work is to confirm the presence of GH{\sc ii}Rs in galaxies visible from the southern hemisphere and to perform a comparative study of GH{\sc ii}Rs with different metallicities and environments.This is performed determining the velocity dispersion which broadens the profile of the emission lines. 
To study the correlation within the Luminosity versus Velocity Dispersion plane, we measure the velocity dispersion from high-resolution spectra as is the echelle spectroscopy. And, using simple dispersion and echelle spectra, we are studying the basic parameters of the regions, such as electronic temperature and density, chemical abundances and evolutionary states of the regions. 

Using narrow-band CCD photometry, Feinstein \cite{F97} analyzed the brightness distribution in a sample of spiral galaxies visible in the southern sky. The most luminous HII regions of Feinstein's sample was our first candidates to study.
In Firpo et al. \cite{Firpo05} and Firpo et al. \cite{2010MNRAS.406.1094F} we confirmed the giant nature of nine candidates to giant HII regions. Continuing our detailed analysis of GHIIRs in local universe galaxies, our sample now includes six spirals galaxies. From photometry of emission lines (Cair\'os et al.\cite{2001ApJS..133..321C}, Gil de Paz et al.\cite{2003ApJS..147...29G}), we have selected bright knots in low-metallicity galaxies, as blue compact dwarf galaxies (BCDs). The bright knots are the new candidates to study.

\section{Observations and Reductions}

We obtained high dispersion spectra in five knots of Haro 15 galaxy with an echelle spectrograph at the 100-inch du Pont Telescope, Las Campanas Observatory (LCO) between 19 and 20, July 2006. Following the same nomenclature of Cair\'os et al.\cite{2001ApJS..136..393C} labelling the bright galactic optical center as A, B is the bright region at the south-east of the galactic center, a weak region C in the north-east, and two new diffuse regions E at the north-northeast close to knot C, and region F at the south-west of the galactic center (E and F regions not are referred by Cair\'os et al. \cite{2001ApJS..136..393C}). The spectral range covered was from 3400 to 10000\AA\ with $\Delta\lambda$=0.25\AA\ at $\lambda$ 6000\AA, as measured from the FWHM of the ThAr comparison lines. This translates in a resolution of $\sim$12 kms$^{-1}$ (R$\simeq$25000). We also obtained long-slit low resolution spectra using the Wide-Field CCD (WFCCD) camera at the same telescope (2005 September 28) for two luminous knots in the BCD galaxy Haro 15. The TEK5 detector was used to cover the wavelength range 3800-9300\AA\ (centered at $\lambda_{c}$ 3800\AA) giving a spectral dispersion of 3.2\AA\ px$^{-1}$ (R$\simeq800$). 
Spectrophotometric standards, according to the respective observing mode, were also observed. We have obtained a good flux calibration for each group of data.

The data analysis was carried out with IRAF\footnote{Image Reduction and Analysis Facility, distributed by NOAO, operated by AURA, Inc., under agreement with NSF.} software. After bias subtraction and flat field corrections with Milky Flats, the bidimensional images were corrected for cosmic rays and reduced with IRAF routines following similar procedures to those described in Firpo et al. \cite{Firpo05}.

\section{The nature of giant HII regions}

In Firpo et al. \cite{2010MNRAS.406.1094F} we observed a residual present in the wings of several lines when fitting single Gaussian profiles to the emission lines.  
Basing on the variety studies that have been proposed in the literature to interpret the existence of the broad supersonic component measured in the emission line profile of GH{\sc ii}Rs, and whenever possible, we have evaluated the possible presence of a broad component (Mu\~noz-Tu\~n\'on et al. \cite{MT96}, Melnick et al.\cite{M99}, H\"agele et al.\cite{H07}, H\"agele et al.\cite{2009MNRAS.396.2295H}, H\"agele et al.\cite{Hagele+10}, among others) or two symmetric low-intensity components in the fit with the observed emission line profile widths (Chu \& Kennicutt \cite{CHK94}, Rala\~no et al. \cite{Relano05}, Rozas et al. \cite{2006A&A...455..539R}).  

As already reported in Firpo et al. \cite{2010MNRAS.406.1094F}, in the present work we have also found that all Haro 15 knots show evidence of wing broadening evident mainly in the H$\alpha$ line and confirmed in other emission lines. Making use of the iterative fitting of multiple Gaussian profiles we evaluate the presence of a broad component and more than one narrow component in the emission line profileAnd, in this case, we have been able to fit a low amplitude broad component to the integral profile wings for all regions.

\begin{figure}
\begin{center}
\includegraphics[width=10cm,angle=0,clip=true]{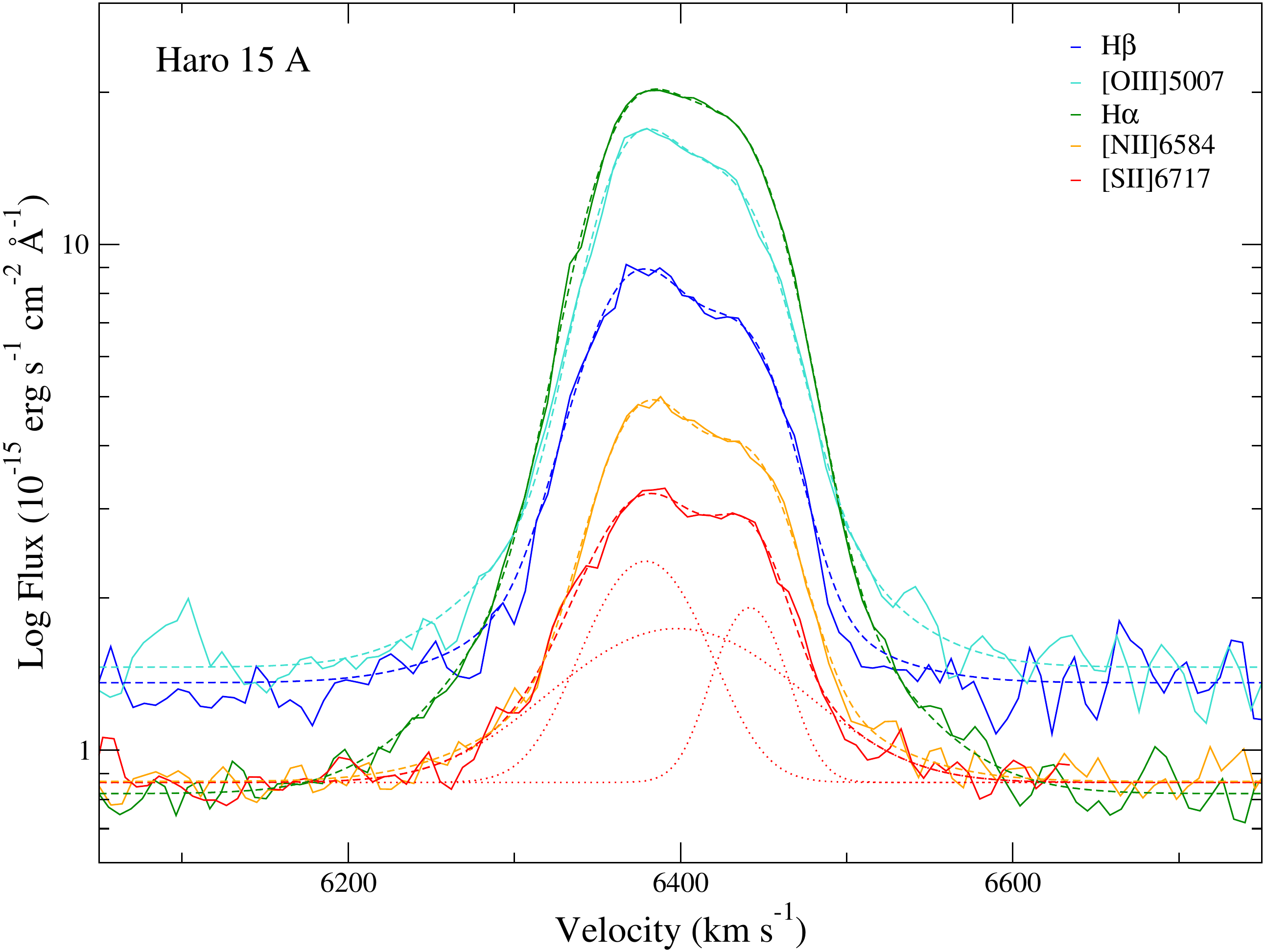}
\caption{\label{figVHA}The agreement between fits to the most intense emission lines can be readily seen in the $\log$(Flux) vs. Velocity plane.}
\end{center}
\end{figure}

Figure~\ref{figVHA} shows the good correlation of the fits between the emission line profiles in the Log Flux-Velocity plane for the most intense emission lines of Knot A.

\section{Relation between H$\alpha$ Luminosities and Velocity Dispersion} 

\begin{figure}
\begin{center}
\vspace{1.0cm}
\includegraphics[width=10cm,angle=0,clip=true]{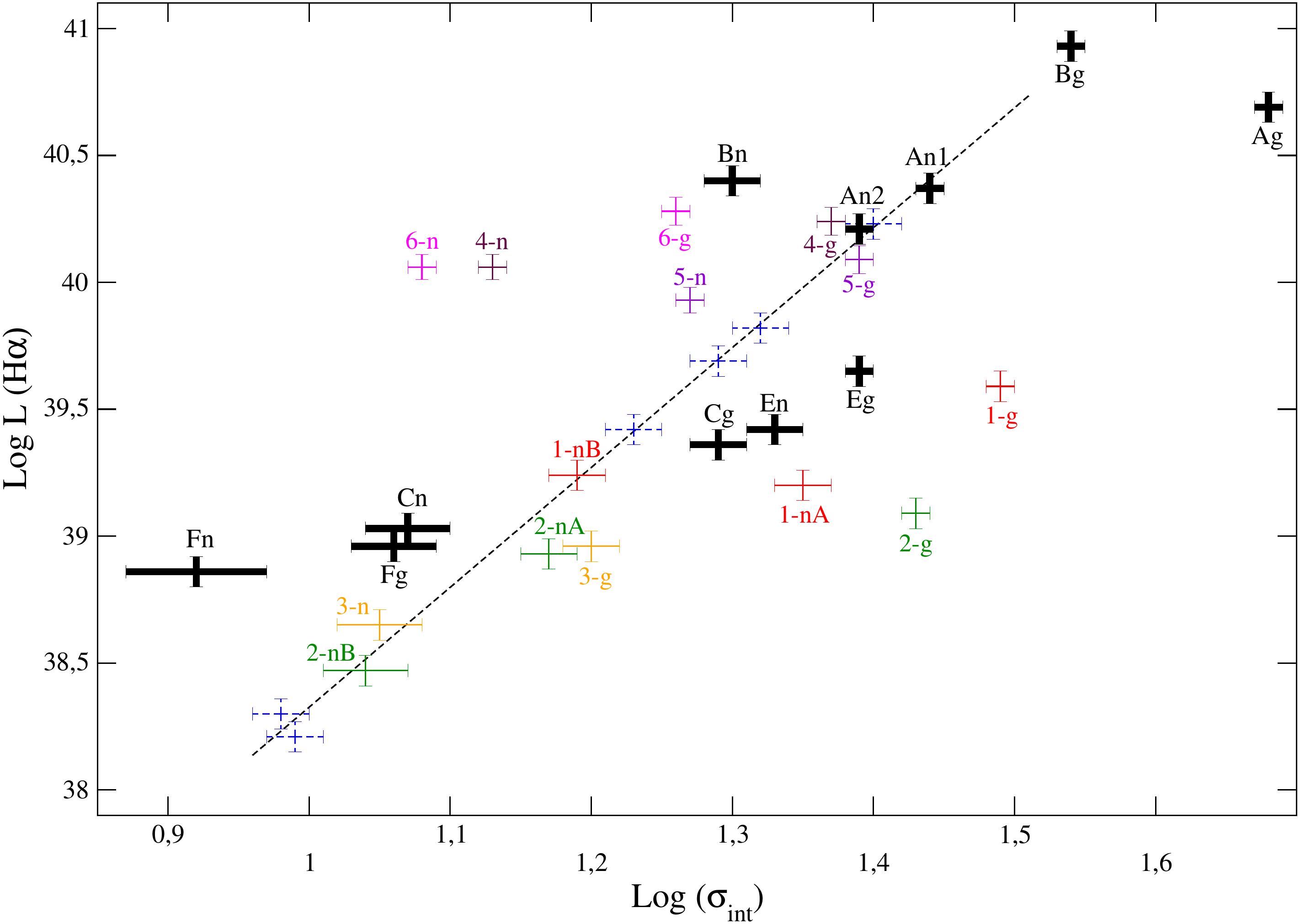}
\caption{\label{Fig:elesigma2}$\log(L) - \log(\sigma)$ relation for our HII regions. Luminosities and velocity dispersions are derived from our spectrophotometric data. The luminosities are not corrected for extinction.}
\end{center}
\end{figure}

The distribution of the regions in the $\log(L) - \log(\sigma)$ plane are strongly dependent on the components derived from the profile fitting (Firpo et al. \cite{2010MNRAS.406.1094F}). Individual components have smaller fluxes and velocity dispersions than the global profile and points are therefore shifted in the diagram. All of the observed knots, except knot F, show supersonic velocities ($>$13~km\,s$^{-1}$).
Figure~\ref{Fig:elesigma2} shows the L(H$\alpha$) vs. $\sigma$ relation for the individual knots of Haro 15 wich are identified from A to F letters. The plot includes results from individual components (black plus pattern symbols with the error bars). Narrow components are identified with subscript n while subscript s refers to line widths measured by fitting a single Gaussian component to the line. H$\alpha$ luminosities were derived from the fluxes measured directly from the component fitting to our echelle spectra (uncorrected by reddening), and using distances as published by de Vaucouleurs et al. \cite{RC3.9} for Haro 15. Data for NGC\,6070 and NGC\,7479 from Firpo et al. \cite{2010MNRAS.406.1094F} (narrow, nA and nB where applicable, and single (g) components) are also plotted and identified by color solid error bars and with numbers (from 1 to 6): NGC\,7479-I (1) in red, NGC\,7479-II (2) in green, NGC\,7479-III (3) in yellow, NGC\,6070-I (4) in maroon, NGC\,6070-II (5) in violet and NGC\,6070-IV (6) in magenta). And a few Giant H{\sc ii} regions from Bosch et al. \cite{B02} are plotted too (blue dashed error bars) together with their linear fit to their ``young'' Giant H{\sc ii} regions as a reference value.
Definitely, the presence of more than one Gaussian component rules the final position of the H{\sc ii} regions in the $\log(L) - \log(\sigma)$ plane. At any case, it can be said that the single Gaussian component represents the upper limit for the velocity dispersion. Characteristic broad components with large supersonic velocity dispersions and low intensities, should not contribute substantially to the total luminosity but could contribute to the observed velocity dispersion.

\section{Physical properties in Haro 15 knots}

From long-slit and echelle spectroscopy, and based on the kinematic results, we study the physical conditions (electron density and temperature), ionic and total chemical abundances of several atoms, reddening and ionization structure derived for each component. 

It was only possible to derive the [O{\sc iii}] and [S{\sc iii}] temperature for knot B, since we are able to measure the corresponding auroral lines only for this source. Temperatureas derived using the direct method from the broad and narrow components are very similar, within the errors.

Oxygen abundances and their uncertainties were derived for each observed knot using the direct method, where possible, or several empirical methods using the strong emission lines present in the spectra. We notice a difference in the O/H ratio between knots A and B. This difference was suggested by L\'opez-S\'anchez \& Esteban \cite{2009A&A...508..615L} as the two objects might have had a different chemical evolution. Our results support these differences between the two regions. Knot C shows oxygen abundance similar to that of knot B within the errors, while the oxygen abundance derived for knot E is closer to the abundance calculated for knot A. The S/N in our knot F spectra is not as good as for the other knots, and the quantities derived for this region should be used with caution.  

The ratio between O$^+$/O$^{2+}$ and S$^+$/S$^{2+}$ denoted by $\eta$ is intrinsically related to the shape of the ionizing continuum and depends on nebula geometry only slightly (V\'{\i}lchez \& Pagel \cite{1988MNRAS.231..257V}). The purely observational counterpart, the $\eta$' diagram ($\eta$'=[([O{\sc ii}]/[O{\sc iii}])/([S{\sc ii}]/[S{\sc iii}])), where $\eta$ and $\eta$' are related through the electron temperature but very weakly. The position of knot B in both diagrams shows a compatible ionization structure, lying in the highest excitation region, similar to the  ionization structure as H{\sc ii} galaxies. The efective temperature of the radiation field (related to the slope of the line) derived for each component in knot B is practically the same. 

\section{Integral Field Spectroscopy in Haro 15 knots}

The net effect on the overall spectrum for each region is to increase the apparent width of the line profile, which could lead to an overestimation of its velocity dispersion as well as a global and not discrete estimation of the physical condition and element abundances of the gas in those regions.

Knots A and B show complex structure which is evident in radial velocity space, but could not be spatially resolved. Recently in 2008, we have obtained high spatial resolution spectroscopy using the Integral Field Unit mode on the Gemini Multi-Object Spectrograph instrument at the Gemini South telescope. The observations were done using the smaller field of view (maximum wavelength coverage) combining the blue grating and the red grating to cover the blue-to-red end. This allowed us to obtain full coverage from [O{\sc iii}]$\lambda$4363\AA\ in the blue to [S{\sc iii}]$\lambda$9532\AA\ in the red. We obtained three observing positions, two are included to observe the regions that build knot A and a offset more is included to observe knot B. To perform an in-depth 2D spectroscopic study, we have obtained the data cubes, and from the preliminar results we are able to resolve the complex kinematical structure within star forming knots in Haro 15 galaxy. 
%
%
%
\small  
%
\section*{Acknowledgments}   
%
We are grateful to the director and staff of LCO for technical assistance and warm hospitality.
This research has made use of the NASA/IPAC Extragalactic Database (NED) which is operated by the Jet Propulsion Laboratory, California Institute of Technology, under contract with the National Aeronautics and Space
Administration.  
Support from the Spanish MEC through grant AYA2007-67965-C03-03 and from the Comunidad de Madrid under grant S-0505/ESP/000237 (ASTROCAM) is acknowledged by GH. VF and GB thank the Universidad Aut\'onoma de Madrid, specially to \'Angeles D\'{\i}az, for their hospitality.
%

%
\end{document}